%% file: uaotim.tex
\documentstyle[psfig]{mn}

\footnotesize
\newdimen\digitwidth    
\setbox0=\hbox{\rm0}
\digitwidth=\wd0
\catcode`!=\active
\def!{\kern\digitwidth}
\normalsize

\newcommand\nuddd{\ifmmode\stackrel{\bf \,...}{\textstyle \nu}\else$\stackrel{\,...}{\textstyle \nu}$\fi}
\def\lsim{~\rlap{$<$}{\lower 1.0ex\hbox{$\sim$}}}

\title{Pulsar Timing at Urumqi Astronomical Observatory:
Observing System and Results}

\author[Wang, Manchester, Zhang et al.]{N. Wang,$^{1,2,3}$\thanks{Email: wangna@ms.xjb.ac.cn}
~~R. N. Manchester,$^{2}$  ~~J. Zhang,$^{1}$   ~~X. J. Wu,$^{3}$   
\newauthor
A. Yusup,$^{1}$  ~~A. G. Lyne,$^{4}$  ~~K. S. Cheng$^{5}$ and ~~M. Z. Chen$^{1}$  \\     
$^{1}$ Urumqi Astronomical Observatory, NAO-CAS, 40 South
Beijing Road, Urumqi, 830011, China\\  
$^{2}$ Australia Telescope National Facility, CSIRO, PO Box
76, Epping, NSW 1710, Australia\\
$^{3}$ Astronomy Department, Peking University, Beijing,
100871, China\\ 
$^{4}$ University of Manchester, Jodrell Bank Observatory,
Macclesfield, Cheshire SK11 9DL\\
$^{5}$ Physics Department, Hong Kong University}

\begin{document}
\maketitle
\pagestyle{plain}

\begin{abstract}
A pulsar timing system has been operating in the 18-cm band at the Urumqi
Astronomical Observatory 25-m telescope since mid-1999. Frequency resolution
allowing dedispersion of the pulsar signals is provided by a
$2\times128\times2.5$ MHz filterbank/digitiser system. Observations of 74
pulsars over more than 12 months have resulted in updated pulsar periods and
period derivatives, as well as improved positions. Comparison with
previous measurements showed that the changes in 
period and period derivative tend to have the same sign and to be correlated
in amplitude. A model based on unseen glitches gives a good explanation of
the observed changes, suggesting that long-term fluctuations in period and
period derivatives are dominated by glitches. In 2000 July, we detected a
glitch of relative amplitude $\Delta\nu/\nu\sim24\times10^{-9}$ in
the Crab pulsar. The post-glitch decay appears similar to other large
Crab glitches. 
\end{abstract}


\section{Introduction}
Pulsars are remarkable celestial clocks. They act as probes for many
different studies, including investigations of the interstellar medium,
stellar and binary system evolution, relativistic astrophysics and condensed
matter. Identified as rotating neutron stars, they have an observed pulse
period range of 0.0016 to 8.5 s. Because of the huge moment of inertia
($\sim 10^{45}$ g cm$^2$) and compact nature of a neutron star, the
rotation periods are very stable. However, they are not completely constant,
increasing gradually as the pulsar loses energy in the form of relativistic
particles and electromagnetic radiation with small variations due to changes
in the stellar structure.  Pulse timing observations can be used to
accurately measure the pulse periods and these variations, including effects
related to the orbital motion for binary pulsars, precise positions and
proper motions and even annual parallax for some pulsars.

If we express the pulsar rotation frequency as a Taylor series, the
pulse phase at time $t$ is given by 
\begin{equation}
\phi(t)=\phi_0+\nu t+\frac{1}{2}\dot\nu t^2+\frac{1}{6}\ddot\nu
t^3+\cdots, 
\label{eq:phase}
\end{equation}
where the rotation frequency $\nu=1/P$, where $P$ is the pulse period,
$\dot\nu$ and $\ddot\nu$ are the first and second time derivatives of
the pulse frequency, and $t$ is the time in a frame which is inertial
with respect to the pulsar, normally taken to be the solar-system
barycenter frame.  The differences between observed pulse arrival
times (TOAs) and those predicted from Equation~\ref{eq:phase}, are
known as timing residuals. Timing residuals will deviate from zero if
the model is not accurate, for example, due to errors in the frequency
parameters or irregularities in the pulse period. Given a set of TOAs
from timing observations at different epochs $t$, corrections to the
rotation and astrometric parameters can be obtained by performing a
least-square fit to minimise the residuals. Improved positions, with
an accuracy of the order 0.1 arcsec, can be obtained from TOAs
spanning a year or more, because an error in position gives rise to an
annual term in the timing residuals due to the orbital motion of the
Earth. 

Two kinds of timing irregularities are observed in pulsars.  The first is
timing noise, characterised by random period changes with the
fractional amplitude about $10^{-9}$ s or less with time scales of
days, months or years \cite{cd85,antt94}. The second kind of timing
irregularity is known as a `glitch', where 
the pulse frequency  suddenly increases  with  the
fractional amplitude $\Delta\nu/\nu$ typically in the range $10^{-9}$ --
$10^{-6}$ \cite{lsg00,wmp+00}. Glitches are 
unpredictable but typically occur at intervals of a few years in young
pulsars. Coincident with the glitch, there is often an increase in the
magnitude of the frequency derivative, typically by $\sim 1$ per cent,
which sometimes has an approximately exponential decay with a time scale of
weeks to years. 

Because of these unpredictable changes, period parameters lose their
validity with time.  However, for some pulsars, parameters have not
been updated for 30 years \cite{tml93}, and hence it is desirable to
improve them. Moreover, to study the properties of timing noise and
glitches, frequent timing observations are required.

The 25-m radio telescope at Nanshan, operated by Urumqi Astronomical
Observatory (UAO), is ideal for such applications. It is mainly used as
a VLBI station and has receivers for six wavelength bands: 1.3~cm,
3.6/13~cm, 6~cm, 18~cm and 92~cm. The telescope is available for
pulsar timing for about two days every week, and so frequent
observations are possible. To overcome the undesirable effects of high
Galactic background emission and of propagation in the interstellar
medium, especially scattering at longer wavelength, we 
decided to use the 18-cm band for the pulsar timing observations. A timing
system based on a room-temperature receiver has been operational since
mid-1999. In Section~\ref{sec:rec} of this paper, the receiver and
data acquisition system are introduced. Observations and data analysis
are discussed in Section~\ref{sec:obs}. In Section~\ref{sec:res}, we
present new measurements of positions, discuss period and period
derivative  changes and introduce 
a glitch detected in the Crab pulsar.  We discuss and summarise our
work in Section~\ref{sec:dis}.         

\section{Receiver and Data Acquisition System}\label{sec:rec}
\begin{figure}
\centerline{\psfig{file=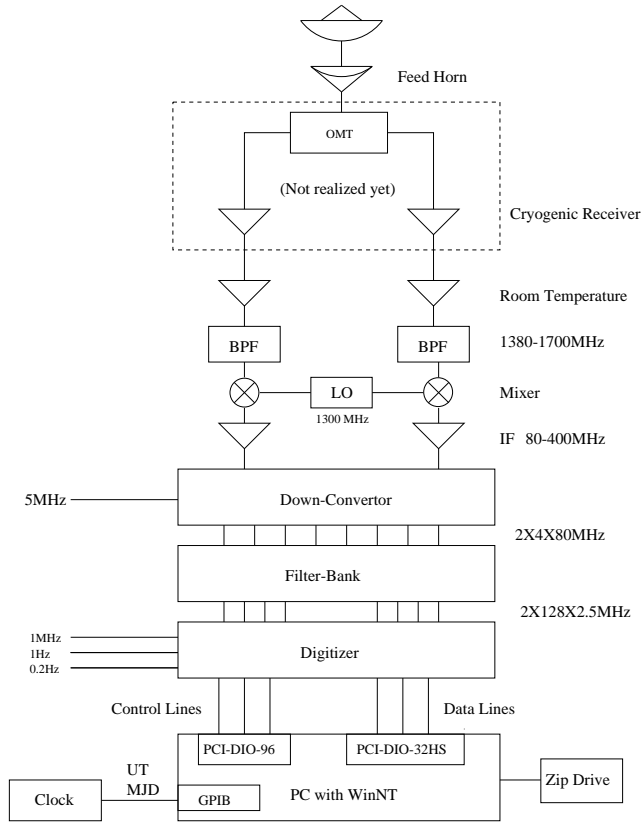,width=85mm}} 
\caption{Diagram of the pulsar timing system at Urumqi Astronomical
observatory. The center frequency is 1540 MHz and the received
bandwidth and the received band is defined by bandpass filters (BPF)
after the RF amplifiers. The ortho-mode transducer (OMT) and the
receiver preamplifiers are currently at room temperature but will soon
be replaced by a cryogenically cooled system. De-dispersion is
provided by a $2\times128\times2.5$ MHz filterbank and digitiser. The
digitiser is controlled by a PC computer using National Instruments
interface cards.}
\label{fg:uaosys}
\end{figure}

A pulsar timing system has been established at the Nanshan Station of
Urumqi Astronomical Observatory, located near the geographic center of
Asia, with longitude $87\degr$ and latitude $+43\degr$. The timing
system operates in the 18-cm band and is outlined in
Fig.~\ref{fg:uaosys}. For this band, the telescope has cassegrain
optics and uses a horn feed receiving orthogonal circular
polarisations. The receiver has dual-channel, room-temperature
pre-amplifiers with center radio frequency (RF) of 1540 MHz and  
total bandwidth of 320 MHz. The system temperature is approximately 100
K.  The polarisations are  split by  an ortho-mode transducer (OMT) at
the end of the feed, amplified and then  down-converted to
intermediate frequency (IF) in the range 80--400 MHz using  a local
oscillator (LO) at 1300 MHz. After conversion, the signals are fed
to a filterbank system which has 128 2.5-MHz channels for each
polarisation.   

Each of the 256 analogue signals from the filterbank pass through a 0.2 Hz
high-pass filter and are then integrated and 1-bit digitised. Digitised data
are grouped in blocks of 32 bits for transfer to the on-line computer.  The
digitiser uses 1-sec and 5-sec pulses derived from the Observatory H-maser
to start the data conversion at a known time. The observation is started by
the 1-sec pulse and the start time is checked using the 5-sec pulse.  The
sampling interval is adjustable as a multiple of 1~$\mu$s, clocked by a
1~MHz reference signal, also derived from the Observatory H-maser. The first
word of each sample is a counter which is checked by the data acquisition
program. A GPS time-transfer system is used to align the Observatory clock
with UTC.

\begin{figure}
\centerline{\psfig{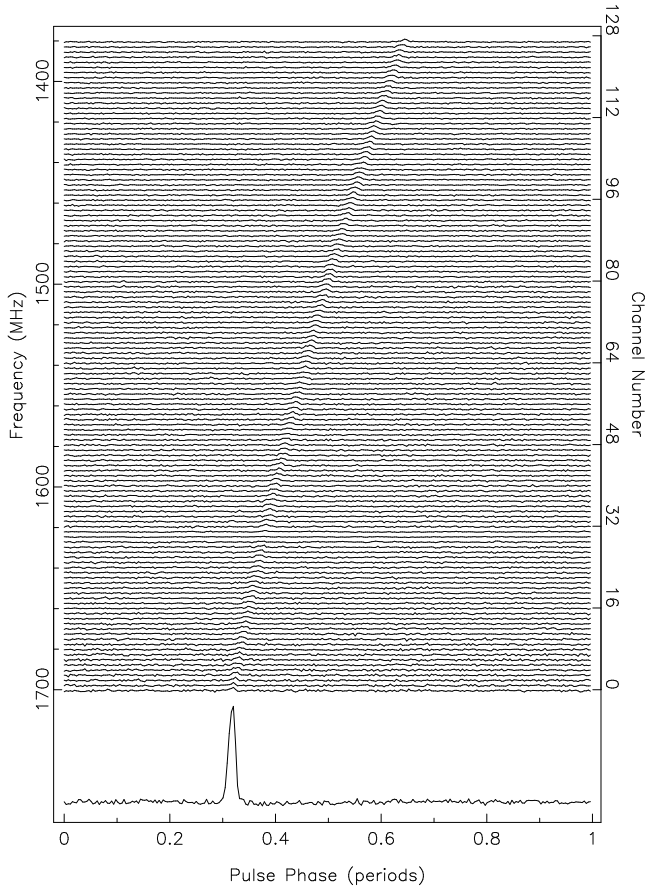}} 
\caption{Pulse profiles of PSR B1933+16. The upper part shows the
profile in each of the 128 2.5MHz channels, the lower part is the sum
of all the channels after appropriate shifts for dispersion have been
applied.}
\label{fg:1933}
\end{figure}

Signals from each channel are recorded by a data acquisition system
based on a PC operating under Windows NT. The online program (Pulsar
Timing Data Acquisition) is written in Visual C++. It has threads to
perform several tasks at the same time, including setting the
digitiser parameters and data sampling, unpacking data and checking time
synchronisation, folding of channel data and saving data to disk,
dedispersing and displaying of folded profiles and communicating with
the computer which controls the telescope. The program reads in the
date and UTC from the clock. A run-line is set high before the chosen
start time and the sampling is triggered by the rising edge of the
next 1-sec pulse. Unpacked data for each channel are folded at the
apparent pulsar period, updated at 120-s intervals, to form
sub-integrations of typical duration 240~s. Predicted topocentric
periods are calculated using `polyco' files produced by the pulsar
timing program TEMPO\footnote{See
http://www.atnf.csiro.au/research/pulsar/timing/tempo/}. Folded data
are saved to disk at every sub-integration time. The sequence counter
and synchronisation flag produced by the digitiser are checked every
sample and on each 5 second respectively.  Folded profiles, including
each channel, the sum of each polarisation and total power, are
displayed for monitor purposes (See Fig.~\ref{fg:1933}) and are
refreshed about once per second.

\section{Observations and Data Analysis}\label{sec:obs}
The sensitivity of pulsar observations is affected by  receiver, spillover  and
Galactic background  noise ($T_{\rm rec}$, $T_{\rm spl}$ and $T_{\rm
bkg}$), antenna gain (G), receiver bandwidth ($\Delta\nu$),
integration time ($t_{\rm int}$), effective pulse width (W) and pulse
period (P). The limiting mean flux density  is given by: 

\begin{equation}
S_{\rm min} = \frac{\alpha \beta (T_{\rm bkg}+T_{\rm spl}+T_{\rm
rec})}{G\sqrt{N_{\rm p}\Delta\nu t_{\rm int}}}\sqrt{\frac{W}{P-W}},  
\label{eq:sensitivity}
\end{equation}
where $\alpha\sim5$ is the threshold signal-to-noise ratio for timing,
$\beta\sim1.5$ is digitisation and other processing losses, $T_{\rm
bkg}+T_{\rm spl}+T_{\rm rec}\sim100$~K, $G\sim0.1$~K~Jy$^{-1}$,
$N_{\rm p}=2$  is the number of polarisations and  
$\Delta\nu=320$~MHz \cite{dtws85}. For a typical pulsar with a duty
cycle $W/P\sim 0.05$, Equation~\ref{eq:sensitivity} gives a
sensitivity  of $S_{\rm min}\sim 2.2$~mJy for $t_{\rm int}\sim16$ mins.      

\begin{figure}
\centerline{\psfig{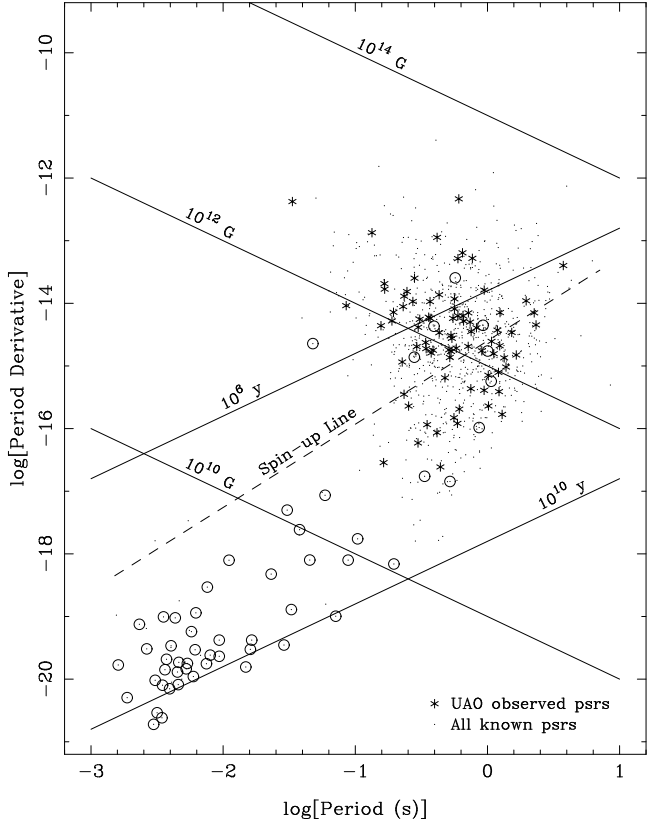}} 
\caption{$P-\dot P$ diagram of known pulsars. Pulsars observed at UAO are
marked as stars. Binary pulsars are indicated by a circle around the point
and lines of constant surface dipole magnetic field and characteristic age
are shown. The spin-up line for millisecond pulsars is indicated by a dashed
line.}
\label{fg:uaolist}
\end{figure}

The pulsar catalogue \cite{tml93} gives flux densities, $S$, for most
pulsars at  400~MHz and 1400~MHz. We first selected pulsars that
are north of declination $-40\degr$ and have $S_{1400}>S_{\rm min}$. 
For those where there are no 1400 MHz measurements, we assume a 
power law spectrum $S\propto f^{\alpha}$, where $f$ is the
observing frequency and $\alpha=-1.7$, to estimate $S_{1400}$ from
$S_{400}$.  All these pulsars were observed from October 1999 to
December 1999. From December, only those pulsars where significant
emission was detected in 16 minutes were observed. There were 74 such
pulsars and these generally had $S_{1400}>4$~mJy. All of these pulsars
are `normal' (non-millisecond) pulsars. Their positions on a $P-\dot
P$  diagram are shown  in Fig.~\ref{fg:uaolist}. Observations are
made once or twice per week with most  being of 16 minutes
duration. In this paper we present observations for these  74 pulsars
up to February 2001.  

\begin{figure}
\centerline{\psfig{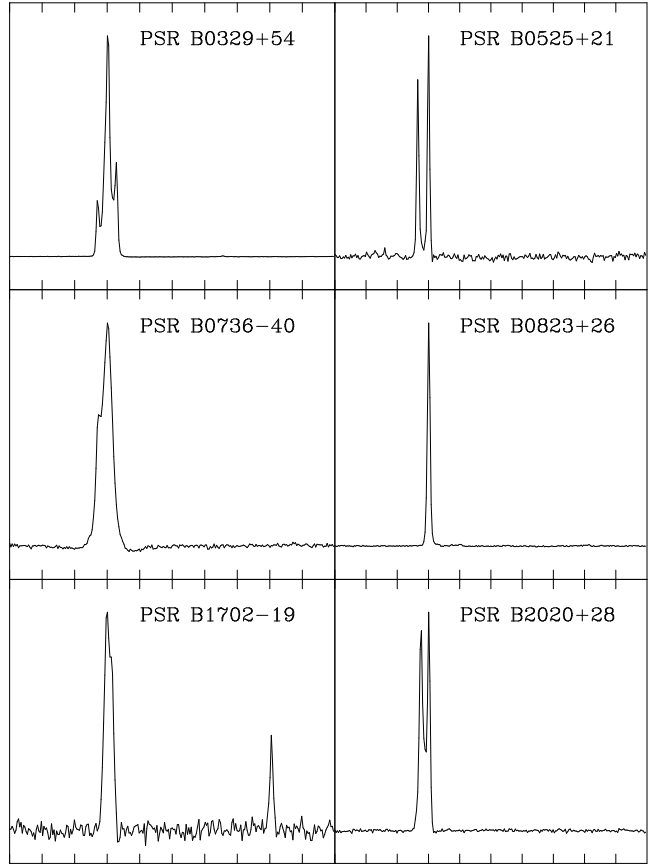}} 
\caption{Examples of standard pulse profiles obtained from the
observations which are used as templates to form the TOAs. The full
pulse period is shown for each profile.}
\label{fg:stdprf}
\end{figure}

After each observation, the data are moved to a Linux system and
compressed using the program TREDUCE (supported by Swinburne
University of Technology and ATNF) to form the archive files. These
archive files consist 8 sub-bands in each polarisation and 4
sub-integrations.  The archives are first dedispersed relative to the
center frequency and summed in time by TREDUCE to form one mean pulse
profile per observation. TOAs are obtained using an iterative
process. First a profile with good signal-to-noise ratio is adopted as
a `standard profile' or template. This template is cross-correlated
with the observed mean pulse profiles to produce TOAs. TEMPO is used
to reduce TOAs to arrival times at infinite frequency at the Solar
system barycentre using the JPL Solar system ephemeris DE200
\cite{sta82}, compute timing residuals and do a least-square fit to
determine a preliminary set of improved model parameters. Using these
parameters, all observed profiles are summed to produce a high
signal-to-noise ratio mean pulse profile which is then adopted as a standard
profile for subsequent TEMPO analyses. As examples,
Fig.~\ref{fg:stdprf} shows six final standard profiles.
For the mode-changing pulsar PSR B0329+54, two standard profiles are
formed for the different modes and TOAs are obtained using the appropriate
profile.

\section{Results}\label{sec:res}
In this section we present the results of timing analyses of 74
pulsars. These results are compared with previously published data
from the pulsar catalogue. In addition we checked recent publications
on pulsar timing and astrometry to obtain the best periods, period
derivatives and positions.  We fit for position, period and period
first derivative for all pulsars except as noted below. The epochs of
periods and positions are fixed at MJD 51700, near the center of our
data span. PSRs B1754$-$24 and B1800$-$21 lie near the ecliptic plane
and so their declinations cannot be determined from timing fits. The
Crab pulsar suffered a glitch during the year and is discussed
separately.  

\subsection{Positions} 
Pulsar positions from the catalogue and this work are present in
Table~\ref{tb:position1}; superscripts `c' and `o' stand for catalogue
value and observed value respectively.  Parameter uncertainties in
this and subsequent tables are given in parentheses and refer to the
last quoted digit. For the observed parameters, they are the
1-$\sigma$ errors given by TEMPO. References for catalogue positions
are given in column 5 and the data span of our observations is given
in the last column.

\input{position1.tex}
\addtocounter{table}{-1}
\input{position2.tex}

Among these pulsars, PSR B1754$-$24 had a very large sinusoid timing
residual before fitting for position. Since it lies close to the ecliptic
plane, timing measurements cannot determine an accurate declination. The
best previously published measurements are R.A.(J2000) = 17$^{\rm
h}$57$^{\rm m}$40$^{\rm s}$(60) \cite{tml93} and Dec.(J2000) =
$-$24$\degr$21$\arcmin$57$\arcsec$(30) \cite{vms83}. In our analysis, we fix
the declination at the catalogue value and fit for R.A.  along with the
other timing parameters. The error of R.A. was estimated by taking the
quadrature sum of the TEMPO R.A. error at the nominal declination and half
of the difference in the R.A. in TEMPO fits with the declination held at
each end of its uncertainty range.  This gives the greatly improved value
R.A.(J2000) = 17$^{\rm h}$57$^{\rm m}$29$\fs$32(3). Similarly, for PSR
B1800$-$21 we obtained R.A.(J2000) = 18$^{\rm h}$03$^{\rm m}$51$\fs$15(3),
significantly different from the catalogue value.

In principle, the parameters given in Table~\ref{tb:position1} can be used
to estimate the pulsar proper motions and (two-dimensional) space
velocities. However, with only one year of data, it is impossible to
separate the effects of proper motion and long-period timing noise, leading
to possible systematic errors in the derived proper motions. We therefore
defer this analysis to a time when more data are available.

\subsection{Period Parameters}\label{subs:period}
Improved periods and period derivatives at epoch MJD 51700 are given in
Table~\ref{tb:rotation1} for all observed pulsars
except the Crab pulsar. Column 5 gives the number of TOAs and  column 6 the
timing residual in microseconds.  

By comparing  the observed and catalogue values, 
we obtain the period and period  derivative changes,  $\Delta P$ and
$\Delta \dot P$ given in columns 7 and 8 respectively. The time
interval from the epoch of the catalogue period to MJD
51700 is given in column 9. There are seven pulsars for which glitches have
been detected in previous observations: PSRs B0355+54, B0525+21,
B0531+21, B1508+55, B1737$-$30, B1800$-$21 and B1830$-$08
\cite{lsg00,wmp+00}.  We quote the latest published post-glitch
rotation parameters  for comparison.  The Crab pulsar (PSR
B0531+21) glitched in July 2000; results for it are discussed in
Section~\ref{subs:crab}.  
 
\input{rotation1.tex}
\addtocounter{table}{-1}
\input{rotation2.tex}

Significant differences between the observed and predicted periods are
observed for most pulsars.  Some pulsars, such as PSRs B0628$-$28, B0844$-$35,
B1737$-$39, B1834$-$10 and B1845$-$01, the period changes are positive
and larger than $10^{-8}$~s. In contrast, PSRs B1700$-$32,
B1706$-$16, B1754$-$24 and B1822$-$09 have periods about $10^{-8}$ s
less than predicted. The observed period
change in PSR B1737$-$30 is huge, with $\Delta P \sim
1.2\times10^{-6}$~s, much larger than for other pulsars.  This pulsar  
glitched  frequently during MJD 46301--49400, with nine glitches
being detected \cite{sl96}, i.e., on average, one glitch every 11
months. It has been more than six years since Shemar \& Lyne's work until
our observations started, so it is probable that there were several
glitches during this interval. 

\begin{figure}
\centerline{\psfig{file=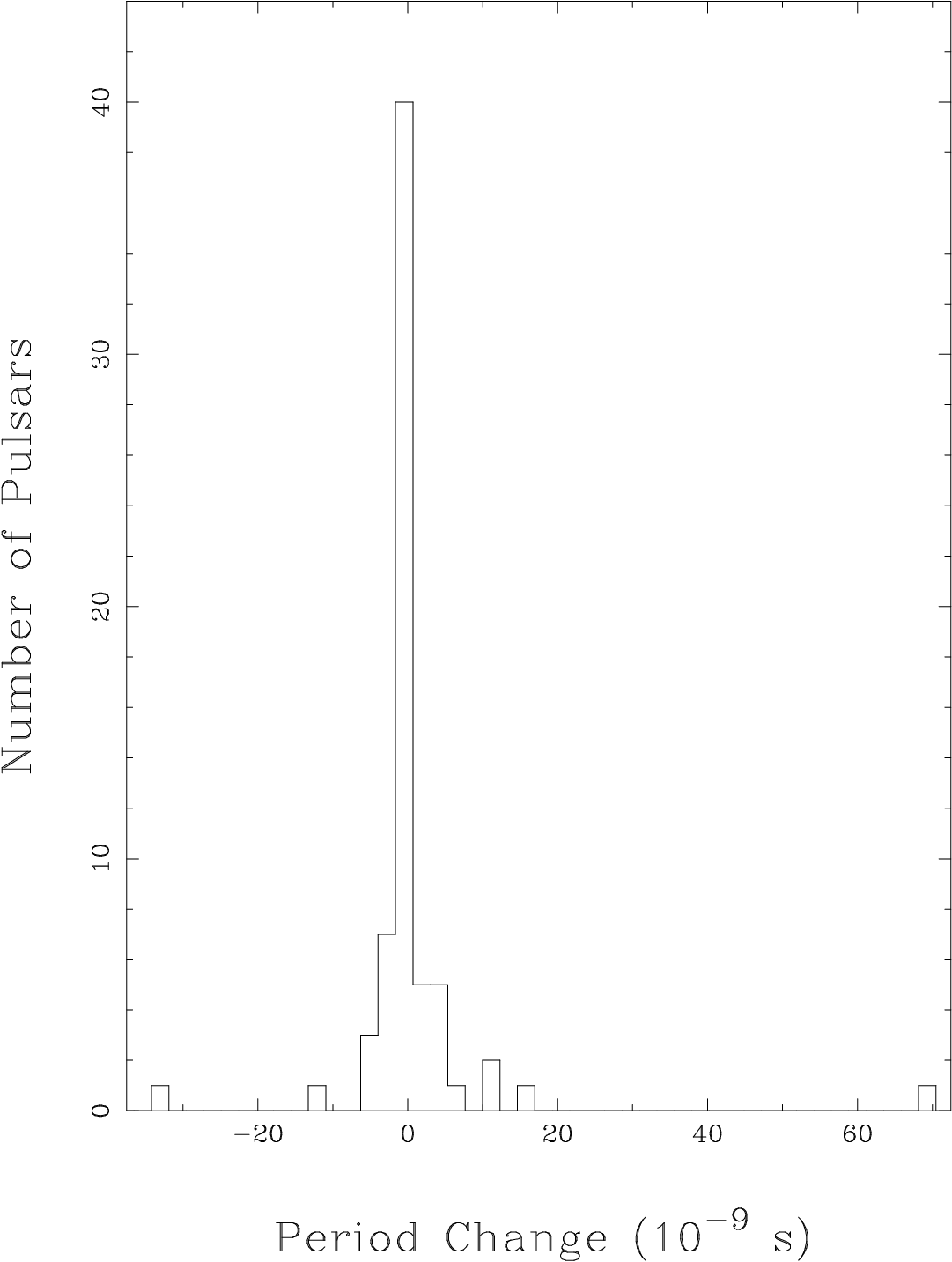,width=70mm}} 
\caption[]{Histogram of pulsar period changes. The Crab pulsar and PSR
B1737$-$30 are not included as they suffered glitches, five other
pulsars with large uncertainties in $\Delta P$ and $\Delta \dot P$ are
excluded as well.}    
\label{fg:dp_hist} 
\end{figure}

Fig.~\ref{fg:dp_hist} shows a histogram of the observed period
changes excepting PSR B1737$-$30 and the Crab pulsar. Also, pulsars are
not included if the uncertainties in $\Delta P$  or $\Delta \dot P$ are
larger than 20~ns or $30\times 10^{-18}$ respectively, eliminating
PSRs B0525+21, B0844$-$35, B1700$-$32, B1754$-$24 and B1800$-$21. With these
exceptions, most period changes are small, with $|\Delta P|\lsim
3$~ns.  

\begin{figure*}
\centerline{\psfig{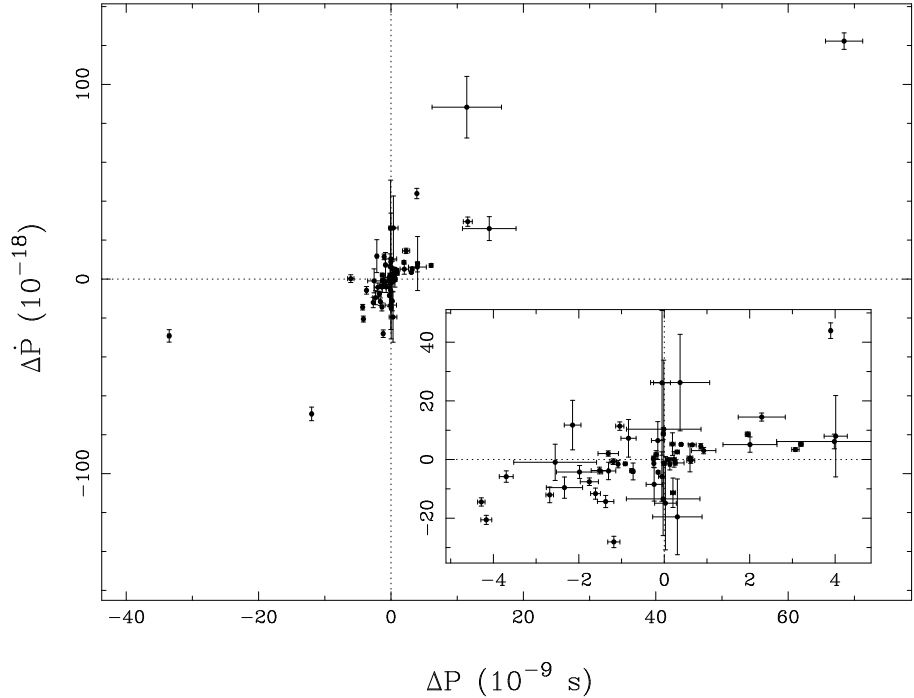}} 
\caption[]{Difference between observed and predicted period
derivative $\Delta \dot P$, plotted against difference in period,
$\Delta P$. The inset is an expanded version of the central
region. The data set is the same as in Fig.~\ref{fg:dp_hist}.}    
\label{fg:dp_dp1} 
\end{figure*}

We find an interesting correlation between $\Delta P$ and $\Delta\dot P$,
illustrated in Fig.~\ref{fg:dp_dp1}. This plot shows a strong tendency for
$\Delta P$ and $\Delta \dot P$ to have the same sign and to be correlated in
amplitude. For random period noise we would expect these two quantities to be
uncorrelated. However, glitches involve systematic changes in both $P$ and
$\dot P$ and we investigate this further.

In most cases, the pulse frequency after a glitch is well modelled by
the relation

\begin{equation}
\nu(t)=\nu_0(t)+\Delta\nu_g[1-Q(1- \exp(-t/\tau_d)] + \Delta\dot\nu_p t,
\label{eq:glmodel}
\end{equation}
where $\nu_0(t)$ is the value of $\nu$ extrapolated from before the glitch,
$\Delta\nu_g$ is the total frequency change at the time of the glitch,
$Q$ is the fractional part of $\Delta\nu_g$ which decays exponentially,
$\tau_d$ is the decay time constant and $\Delta\dot\nu_p$ is the
permanent change in $\dot\nu$ at the time of the glitch
\cite{wmp+00}. The increment in frequency derivative is given by:

\begin{equation}
\Delta\dot\nu(t)=\frac{-Q\Delta\nu_g}{\tau_d}\exp(-t/\tau_d) + \Delta\dot\nu_p.
\label{eq:glnudot}
\end{equation}

\begin{figure}
\centerline{\psfig{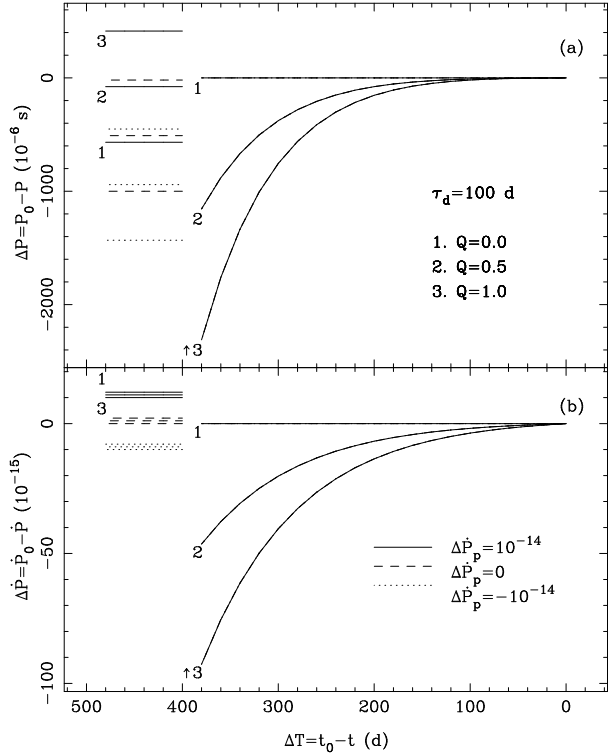}} 
\caption[]{A model for the increments in pulse period and period derivative
due to a glitch occurring 400 days before the current time $t_0$. A glitch
magnitude $\Delta \nu_{\rm g}=10^{-6}$ and a post-glitch decay time of 100
days are assumed. Changes for three different decay parameters $Q$ and three
different permanent changes in period derivative $\Delta \dot P_{\rm p}$ are
plotted. The groups of three lines of one type are for different values of
$Q$ and the solid, dashed and dotted lines represent different values of
$\Delta \dot P_{\rm p}$. The post-glitch values are functions of $Q$ and
independent of $\Delta \dot P_{\rm p}$, so they overlap for different
$\Delta \dot P_{\rm p}$.}
\label{fg:gltmd} 
\end{figure}

We use these relations to compute $P$ and $\dot P$ at an arbitrary time $t$ 
in the {\it past}, and compare them with the values at the {\it
current} time $t_0$, 
assuming a glitch of magnitude $\Delta \nu_{\rm g}=10^{-6}$ at time
$t_0-t=400$ days.  Values of $\Delta P=P_0-P$ and $\Delta \dot P=\dot
P_{0}-\dot P$ are shown as a function of $t_0-t$ in
Fig.~\ref{fg:gltmd}. If a pulsar glitched before the previous observation,
both $\Delta P$ and $\Delta \dot P$ will be negative; while if the glitch
happened between the observations, we have $\Delta P<0$ and $\Delta \dot
P<0$ for negative $\Delta \dot P_{\rm p}=-\Delta \dot\nu_{\rm p}/\nu^{2}$,
and $\Delta P>0$ and $\Delta \dot P>0$ for positive $\Delta \dot P_{\rm p}$
and large fractional decay $Q$. These predictions are consistent with the
observations shown in Fig.~\ref{fg:dp_dp1}. The last case represents the
observed situation in the Crab pulsar \cite{lps93}. For positive $\Delta
P_{\rm p}$ and small $Q$, we can have $\Delta P<0$ and $\Delta \dot
P>0$. Furthermore, in this model there is no case with $\Delta P>0$ and
$\Delta \dot P<0$, consistent with the observed deficit in the lower-right
of Fig.~\ref{fg:dp_dp1}. For $Q>0$ in a glitch before the first
observation, the changes in $\Delta P$ and $\Delta \dot P$ are proportional,
also consistent with the observations shown in Fig.~\ref{fg:dp_dp1}. These
agreements strongly suggest that long-term systematic changes in pulsar
periods and their derivatives are dominated by pulsar glitches (cf. Lyne
1996;\nocite{lyn96} Johnston \& Galloway 1999\nocite{jg99}).
               
\subsection{The Crab Pulsar}\label{subs:crab}
Observations of the Crab pulsar revealed a glitch during 2000 July.
This glitch was also observed at Jodrell Bank Observatory and
the pre-glitch ephemeris given in Table~\ref{tb:eph} is based on
Jodrell Bank data. They also determined the glitch epoch to be MJD
51740.8 (2000 July 15). We use these parameters in our analyses. 
\begin{table}
\caption{Pre-glitch ephemeris of the Crab pulsar.}
\begin{tabular}{ll}
\hline & \vspace{-3mm} \\
PSR         		& B0531+21 (J0534+2200) \\
RAJ        		& 05$^{\rm h}$34$^{\rm m}$31$\fs$972  \\
DECJ        		& 22$\degr$00$\arcmin$52$\farcs$07  \\
PEPOCH      		& 51562.7279   \\
$\nu$ (s$^{-1}$) 	& 29.845547780  \\
$\dot\nu$ (s$^{-2}$) 	& $-$3.7457341$\times10^{-10}$ \\
$\ddot\nu$ (s$^{-3}$) 	& 1.0161006$\times10^{-20}$ \\
$\nuddd$ (s$^{-4}$) 	& $-$6.0$\times10^{-31}$ \\
DM (cm$^{-3}$ pc)   	& 56.77  \\
\hline&\vspace{-3mm} 
\end{tabular}
\label{tb:eph}
\end{table}

\begin{figure} 
\begin{center} 
\begin{tabular}{c} 
\mbox{\psfig{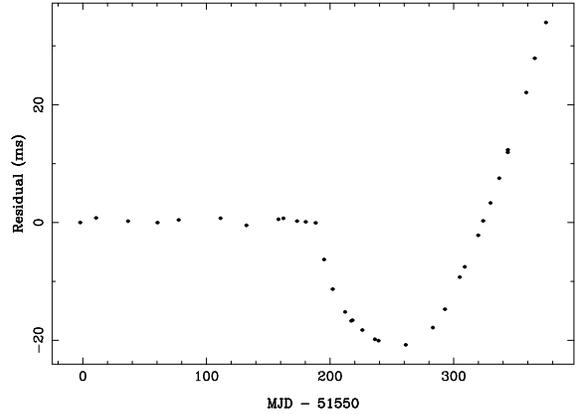}} 
\end{tabular} 
\caption[]{Timing residual of the Crab pulsar showing the glitch of 2000 July.}  
\label{fg:crab_res} 
\end{center} 
\end{figure}

\begin{figure} 
\begin{center} 
\begin{tabular}{c} 
\mbox{\psfig{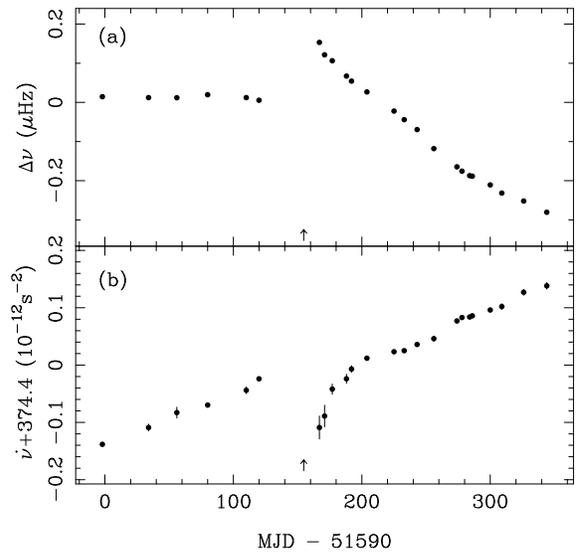}} 
\end{tabular} 
\caption[]{A glitch of the Crab pulsar at epoch MJD 51741. (a) frequency
residual $\Delta\nu$ relative to the pre-glitch solution and (b) the
variation of $\dot\nu$. The glitch epoch is indicated by an arrow near
the bottom of each plot.}
\label{fg:glitch} 
\end{center} 
\end{figure}

Residuals relative to the model in Table~\ref{tb:eph} are shown in
Fig.~\ref{fg:crab_res} and Fig.~\ref{fg:glitch} shows the observed
variations in frequency and frequency derivative. We estimate the time
constant of the glitch, $\tau_d$, to be 4 days. This value gives phase
continuity at the adopted glitch epoch and we keep it fixed during the
analyses. Fitting of the glitch model described in Equation~\ref{eq:glmodel}
using TEMPO gives a glitch size $\Delta\nu/\nu\sim24(8)\times10^{-9}$, an
increment in $\Delta\dot\nu/\dot\nu\sim 5(2)\times10^{-3}$, and $Q=0.8$,
that is 80 per cent of the jump in frequency decays away on the given
timescale $\tau_d$. As may be seen in Fig.~\ref{fg:glitch}, there is a
persistent increase in the magnitude of the slow-down rate; fitting for this
in TEMPO gives $\Delta\dot\nu_p\sim-48(3)\times 10^{-15}\;{\rm
s}^{-2}$. Rotation parameters
for the post-glitch data obtained using TEMPO are given in
Table~\ref{tb:postglitch}.

\begin{table}
\begin{minipage}{8cm}
\caption{Rotation parameters after the 2000 July glitch.}
\begin{tabular}{ll}
\hline & \vspace{-3mm} \\
Epoch (MJD)             & 51856.0000   \\
Data span               & 51745.2--51966.6     \\
$\nu$     (s$^{-1}$)    & 29.836059670(2)   \\
$\dot\nu$ (s$^{-2}$)    & $-3.743460(3)\times10^{-10}$ \\
$\ddot\nu$ (s$^{-3}$)   & $1.17(2)\times10^{-20}$ \\
\hline&\vspace{-3mm} 
\end{tabular}
\label{tb:postglitch}
\end{minipage}
\end{table}

Glitches in the Crab pulsar are frequent and have relative sizes in the range
$4.7\times10^{-9}$ to $85\times10^{-9}$ \cite{lps93,wbl01}. Lyne et
al. (1993) showed that two of the largest Crab glitches ($\Delta\nu/\nu\sim
37$ and $85\times10^{-9}$) had persistent increases in slow-down
rate, so that the Crab pulsar is now rotating more slowly than it would
have without the glitches. It is clear that the 2000 July glitch falls into
the class of large glitches in the Crab pulsar. 

\section{Discussion}\label{sec:dis}
We have developed a pulsar timing system for the 25-m Nanshan radio
telescope of Urumqi Astronomical Observatory and used it to determine 
positions and period parameters for 74 pulsars. Typical timing
residuals over one year are few hundred microseconds, giving positions
typically to a few tenths of an arcsec and pulse periods to 0.01 ns or
better. Comparison with earlier measurements has shown that
long-term period fluctuations are probably dominated by recovery from
(mostly unseen) glitches.

A cryogenic receiver system operating in the 18-cm band for the UAO Nanshan
telescope is currently under construction. This will enable timing and other
measurements to be made on pulsars with mean flux density as low as 1
mJy. Compared to larger radio telescopes, frequent pulsar observations are
possible with this system, making it ideal for projects such as the study of
pulsar timing irregularities and scintillation. The higher sensitivity of
the cryogenic system will allow more precise timing and observations of a
larger sample of pulsars.

\section*{ACKNOWLEDGEMENTS}
We thank the engineers responsible for maintaining
the receiver,  telescope and H-maser at UAO, and the  staff 
who helped with the observations. The filterbank/digitiser system was
designed and constructed at Jodrell Bank Observatory, University of
Manchester, and the downconvertor system was built at the Australia Telescope
National Facility. We thank them for their support.  NW thanks
Beijing Astrophysics Center for providing computer facilities, the
ATNF for a Postgraduate Research Scholarship and the National Nature Science
Foundation of China for supporting the timing project.
 

\end{document}

%% file: position1.tex
\begin{table*}
\begin{minipage}{170mm}
\caption{Pulsars positions from catalogue and this work.}
\begin{tabular}{llllcllc}
\hline &\vspace{-3mm} \\
~~~PSR J                    & ~~PSR B                    &
 !$\alpha^c$(J2000)         & !!$\delta^c$(J2000)          & 
 Ref.\footnote{
   1. Downs \& Reichley, 1983 \nocite{dr83}
~~~2. Arzoumanian et al., 1994 \nocite{antt94}
~~~3. Fomalont et al., 1992 \nocite{fgl+92}
~~~4. Fomalont et al., 1997 \nocite{fgml97}
~~~5. Dewey et al., 1988 \nocite{dtms88}
~~~6. McNamara, 1973, \nocite{mcn71}
~~~7. Siegman, Manchester \& Durdin, 1993 \nocite{smd93}
~~~8. Newton, Manchester \& Cooke, 1981 \nocite{nmc81}
~~~9. Manchester \& Taylor, 1981 \nocite{mt81} 
~~~10. Johnston et al., 1995 \nocite{jml+95}
~~~11. Braun, Goss and Lyne, 1989 \nocite{bgl89}
~~~12. Taylor,  Manchester and Lyne, 1993 \nocite{tml93}
~~~13. Vivekanand, Mohanty \& Salter, 1983 \nocite{vms83}
~~~14. Clifton et al., 1992 \nocite{clj+92}
~~~15. Manchester et al., 1996 \nocite{mld+96}
~~~16. Backus, Taylor \& Damashek, 1982 \nocite{btd82}
~~~17. Gullahorn \& Rankin, 1978 \nocite{gr78}
~~~18. Foster, Backer \& Wolszczan, 1990 \nocite{fbw90}
}  &
 !$\alpha^o$(J2000)         & !!$\delta^o$(J2000)          &
Data Span\\
                            &                            & 
~($^{\rm h}$ ~$^{\rm m}$ ~$^{\rm s}$) & ~~~($\degr$ ~~$\arcmin$ ~~$\arcsec$) &
                            & 
~($^{\rm h}$ ~$^{\rm m}$ ~$^{\rm s}$) & ~~~($\degr$ ~~$\arcmin$ ~~$\arcsec$) &    MJD  \\ 
\hline &\vspace{-3mm} \\
 0034$-$0721 &    0031$-$07 & 00:34:08.88(3)~ &  $-$07:21:53.4(7)~ &     1 & 00:34:08.87(5)~~~~ & $-$07:21:54.7(11)~ &    51560--51948 \\ 
   0139+5814 &      0136+57 & 01:39:19.770(3) &  +58:14:31.85(3)~~ &     2 & 01:39:19.740(7)~~~ & +58:14:31.94(4)~~~ &    51547--51966 \\ 
   0141+6009 &      0138+59 & 01:41:39.947(7) &  +60:09:32.28(5)~~ &     3 & 01:41:39.97(5)~~~~ & +60:09:32.28(17)~~ &    51560--51915 \\ 
   0332+5434 &      0329+54 & 03:32:59.35(1)~ &  +54:34:43.2(1)~~~ &     1 & 03:32:59.367(7)~~~ & +54:34:43.44(9)~~~ &    51499--51967 \\ 
   0358+5413 &      0355+54 & 03:58:53.705(4) &  +54:13:13.58(3)~~ &     4 & 03:58:53.708(6)~~~ & +54:13:13.80(8)~~~ &    51499--51966 \\ 
 \vspace{-2mm} \\ 
 0452$-$1759 &    0450$-$18 & 04:52:34.098(3) &  $-$17:59:23.54(7) &     4 & 04:52:34.096(9)~~~ & $-$17:59:23.48(12) &    51572--51966 \\ 
   0454+5543 &      0450+55 & 04:54:07.621(3) &  +55:43:41.2(1)~~~ &     5 & 04:54:07.724(7)~~~ & +55:43:41.59(9)~~~ &    51547--51966 \\ 
   0528+2200 &      0525+21 & 05:28:52.34(2)~ &  +22:00:00(5)~~~~~ &     1 & 05:28:52.26(3)~~~~ & +22:00:00.5(70)~~~ &    51547--51959 \\ 
   0534+2200 &      0531+21 & 05:34:31.973(5) &  +22:00:52.06(6)~~ &     6 & ~~~~~~~~~~~~~~~~~~ & ~~~~~~~~~~~~~~~~~~ &    51547--51966 \\ 
   0543+2329 &      0540+23 & 05:43:09.650(3) &  +23:29:06.14(4)~~ &     3 & 05:43:09.67(11)~~~ & +23:28:57(49)~~~~~ &    51499--51966 \\ 
 \vspace{-2mm} \\ 
   0612+3721 &      0609+37 & 06:12:48.654(3) &  +37:21:37.0(1)~~~ &     5 & 06:12:48.653(13)~~ & +37:21:36.6(6)~~~~ &    51607--51938 \\ 
 0630$-$2834 &    0628$-$28 & 06:30:49.531(6) &  $-$28:34:43.6(1)~ &     1 & 06:30:49.468(16)~~ & $-$28:34:42.2(3)~~ &    51560--51959 \\ 
 0738$-$4042 &    0736$-$40 & 07:38:32.432(3) &  $-$40:42:41.15(4) &     1 & 07:38:32.313(4)~~~ & $-$40:42:40.25(6)~ &    51586--51966 \\ 
 0742$-$2822 &    0740$-$28 & 07:42:49.073(3) &  $-$28:22:44.0(1)~ &     3 & 07:42:49.030(6)~~~ & $-$28:22:43.25(15) &    51560--51966 \\ 
   0814+7429 &      0809+74 & 08:14:59.44(4)~ &  +74:29:05.8(1)~~~ &     2 & 08:14:59.47(4)~~~~ & +74:29:05.3(4)~~~~ &    51586--51966 \\ 
 \vspace{-2mm} \\ 
 0820$-$1350 &    0818$-$13 & 08:20:26.358(9) &  $-$13:50:55.20(6) &     3 & 08:20:26.372(7)~~~ & $-$13:50:56.0(3)~~ &    51547--51966 \\ 
   0826+2637 &      0823+26 & 08:26:51.310(2) &  +26:37:25.57(7)~~ &     1 & 08:26:51.500(14)~~ & +26:37:25.5(8)~~~~ &    51547--51967 \\ 
 0837$-$4135 &    0835$-$41 & 08:37:21.173(6) &  $-$41:35:14.29(7) &     7 & 08:37:21.184(3)~~~ & $-$41:35:14.36(6)~ &    51547--51966 \\ 
 0846$-$3533 &    0844$-$35 & 08:46:05.9(1)~~ &  $-$35:33:40(2)~~~ &     8 & 08:46:05.85(3)~~~~ & $-$35:33:39.4(10)~ &    51606--51915 \\ 
   0922+0638 &      0919+06 & 09:22:13.977(3) &  +06:38:21.69(4)~~ &     3 & 09:22:13.940(17)~~ & +06:38:18.9(8)~~~~ &    51560--51966 \\ 
 \vspace{-2mm} \\ 
   0953+0755 &      0950+08 & 09:53:09.316(3) &  +07:55:35.60(4)~~ &     3 & 09:53:09.23(3)~~~~ & +07:55:33.2(11)~~~ &    51547--51966 \\ 
   1136+1551 &      1133+16 & 11:36:03.296(4) &  +15:51:00.7(1)~~~ &     9 & 11:36:03.184(17)~~ & +15:51:10.1(5)~~~~ &    51548--51967 \\ 
   1239+2453 &      1237+25 & 12:39:40.475(3) &  +24:53:49.25(3)~~ &     2 & 12:39:40.353(12)~~ & +24:53:50.06(18)~~ &    51548--51966 \\ 
   1509+5531 &      1508+55 & 15:09:25.724(9) &  +55:31:33.01(8)~~ &     2 & 15:09:25.666(12)~~ & +55:31:32.55(6)~~~ &    51548--51966 \\ 
 1645$-$0317 &    1642$-$03 & 16:45:02.045(3) &  $-$03:17:58.4(1)~ &     1 & 16:45:02.043(6)~~~ & $-$03:17:58.3(3)~~ &    51548--51968 \\ 
 \vspace{-2mm} \\ 
 1703$-$3241 &    1700$-$32 & 17:03:22.4(1)~~ &  $-$32:41:45(4)~~~ &     2 & 17:03:22.534(13)~~ & $-$32:41:48.1(6)~~ &    51548--51966 \\ 
 1705$-$1906 &    1702$-$19 & 17:05:36.108(6) &  $-$19:06:38.5(7)~ &     2 & 17:05:36.076(8)~~~ & $-$19:06:36.7(14)~ &    51548--51966 \\ 
 1707$-$4053 &    1703$-$40 & 17:07:21.744(6) &  $-$40:53:55.3(3)~ &    10 & 17:07:21.728(12)~~ & $-$40:53:56.1(4)~~ &    51573--51925 \\ 
 1709$-$1640 &    1706$-$16 & 17:09:26.455(2) &  $-$16:40:58.4(3)~ &     1 & 17:09:26.460(8)~~~ & $-$16:40:58.4(10)~ &    51548--51966 \\ 
 1721$-$3532 &    1718$-$35 & 17:21:32.80(2)~ &  $-$35:32:46.6(9)~ &    10 & 17:21:32.771(11)~~ & $-$35:32:48.6(4)~~ &    51573--51967 \\ 
 \vspace{-2mm} \\ 
 1722$-$3207 &    1718$-$32 & 17:22:02.955(4) &  $-$32:07:44.9(3)~ &     2 & 17:22:02.951(5)~~~ & $-$32:07:45.6(4)~~ &    51506--51967 \\ 
 1740$-$3015 &    1737$-$30 & 17:40:33.7(1)~~ &  $-$30:15:42(2)~~~ &    11 & 17:40:33.753(16)~~ & $-$30:15:43.8(13)~ &    51549--51967 \\ 
 1741$-$3927 &    1737$-$39 & 17:41:18.04(3)~ &  $-$39:27:38(1)~~~ &     8 & 17:41:18.071(6)~~~ & $-$39:27:38.41(20) &    51549--51967 \\ 
 1745$-$3040 &    1742$-$30 & 17:45:56.299(2) &  $-$30:40:23.6(3)~ &     2 & 17:45:56.300(5)~~~ & $-$30:40:23.2(5)~~ &    51549--51967 \\ 
 1752$-$2806 &    1749$-$28 & 17:52:58.66(2)~ &  $-$28:06:48(5)~~~ &     1 & 17:52:58.712(6)~~~ & $-$28:06:35.9(10)~ &    51513--51968 \\ 
 \vspace{-2mm} \\ 
 1757$-$2421 &    1754$-$24 & 17:57:40(60)~~~ &  $-$24:21:57(30)~~ & 12,13 & 17:57:29.328(5)~~~ & ~~~~~~~~~~~~~~~~~~ &     51561--51967 \\ 
 1803$-$2137 &    1800$-$21 & 18:03:51.35(3)~ &  $-$21:37:07.2(5)~ &    11 & 18:03:51.15(4)~~~~ & ~~~~~~~~~~~~~~~~~~ &     51561--51966 \\ 
 1807$-$0847 &    1804$-$08 & 18:07:38.019(9) &  $-$08:47:43.1(2)~ &     3 & 18:07:38.036(5)~~~ & $-$08:47:43.3(4)~~ &    51500--51966 \\ 
 1818$-$1422 &    1815$-$14 & 18:18:23.79(5)~ &  $-$14:22:36(3)~~~ &    14 & 18:18:23.770(5)~~~ & $-$14:22:38.9(5)~~ &    51549--51960 \\ 
 1820$-$0427 &    1818$-$04 & 18:20:52.621(3) &  $-$04:27:38.5(1)~ &     1 & 18:20:52.620(5)~~~ & $-$04:27:37.1(4)~~ &    51506--51967 \\ 
 \vspace{-2mm} \\ 
 1824$-$1945 &    1821$-$19 & 18:24:00.45(4)~ &  $-$19:45:51(8)~~~ &     8 & 18:24:00.449(6)~~~ & $-$19:45:49.9(12)~ &    51512--51966 \\ 
 1825$-$0935 &    1822$-$09 & 18:25:30.596(6) &  $-$09:35:22.8(4)~ &     2 & 18:25:30.599(7)~~~ & $-$09:35:21.9(6)~~ &    51506--51967 \\ 
 1829$-$1751 &    1826$-$17 & 18:29:43.12(1)~ &  $-$17:51:03(2)~~~ &     7 & 18:29:43.130(10)~~ & $-$17:50:57.0(9)~~ &    51588--51966 \\ 
 1832$-$0827 &    1829$-$08 & 18:32:37.024(7) &  $-$08:27:03.7(3)~ &    14 & 18:32:37.017(5)~~~ & $-$08:27:03.0(5)~~ &    51500--51967 \\ 
 1833$-$0827 &    1830$-$08 & 18:33:40.32(2)~ &  $-$08:27:30.7(6)~ &    14 & 18:33:40.302(4)~~~ & $-$08:27:31.50(15) &    51597--51960 \\ 
 \vspace{-2mm} \\ 
 1835$-$1106 &              & 18:35:18.287(2) &  $-$11:06:15.1(2)~ &    15 & 18:35:18.17(3)~~~~ & $-$11:06:19.3(13)~ &    51600--51966 \\ 
 1836$-$1008 &    1834$-$10 & 18:36:53.9(1)~~ &  $-$10:08:09(5)~~~ &    16 & 18:36:53.911(5)~~~ & $-$10:08:09.8(4)~~ &    51500--51966 \\ 
   1840+5640 &      1839+56 & 18:40:44.59(5)~ &  +56:40:55.6(4)~~~ &     2 & 18:40:44.48(4)~~~~ & +56:40:55.0(4)~~~~ &    51548--51966 \\ 
 1847$-$0402 &    1844$-$04 & 18:47:22.83(1)~ &  $-$04:02:14.2(5)~ &     7 & 18:47:22.833(9)~~~ & $-$04:02:13.8(5)~~ &    51550--51966 \\ 
 1848$-$0123 &    1845$-$01 & 18:48:23.60(2)~ &  $-$01:23:58.2(7)~ &     7 & 18:48:23.589(5)~~~ & $-$01:23:57.9(3)~~ &    51500--51966 \\ 
\hline&\vspace{-8mm} 
\end{tabular}
\label{tb:position1}
\end{minipage}
\end{table*}

%% file: position2.tex
\begin{table*}
\begin{minipage}{170mm}
\caption{--{\it Continued.}}
\begin{tabular}{llllcllc}
\hline &\vspace{-3mm} \\
~~~PSR J                    & ~~PSR B                    &
 !$\alpha^c$(J2000)         & !!$\delta^c$(J2000)          & 
  Ref. \footnote{
   1. Downs \& Reichley, 1983 \nocite{dr83}
~~~2. Arzoumanian et al., 1994 \nocite{antt94}
~~~3. Fomalont et al., 1992 \nocite{fgl+92}
~~~4. Fomalont et al., 1997 \nocite{fgml97}
~~~5. Dewey et al., 1988 \nocite{dtms88}
~~~6. McNamara, 1973, \nocite{mcn71}
~~~7. Siegman, Manchester \& Durdin, 1993 \nocite{smd93}
~~~8. Newton, Manchester \& Cooke, 1981 \nocite{nmc81}
~~~9. Manchester \& Taylor, 1981 \nocite{mt81} 
~~~10. Johnston et al., 1995 \nocite{jml+95}
~~~11. Braun, Goss and Lyne, 1989 \nocite{bgl89}
~~~12. Taylor,  Manchester and Lyne, 1993 \nocite{tml93}
~~~13. Vivekanand, Mohanty \& Salter, 1983 \nocite{vms83}
~~~14. Clifton et al., 1992 \nocite{clj+92}
~~~15. Manchester et al., 1996 \nocite{mld+96}
~~~16. Backus, Taylor \& Damashek, 1982 \nocite{btd82}
~~~17. Gullahorn \& Rankin, 1978 \nocite{gr78}
~~~18. Foster, Backer \& Wolszczan, 1990 \nocite{fbw90}
}   &
 !$\alpha^o$(J2000)         & !!$\delta^o$(J2000)          &
Data Span\\
                            &                            & 
~($^{\rm h}$ ~$^{\rm m}$ ~$^{\rm s}$) & ~~~($\degr$ ~~$\arcmin$ ~~$\arcsec$) &
                            & 
~($^{\rm h}$ ~$^{\rm m}$ ~$^{\rm s}$) & ~~~($\degr$ ~~$\arcmin$ ~~$\arcsec$) &    MJD  \\
\hline &\vspace{-3mm} \\
 1900$-$2600 &    1857$-$26 & 19:00:47.60(1)~ &  $-$26:00:43.1(3)~ &     3 & 19:00:47.561(18)~~ & $-$26:00:39(3)~~~~ &    51573--51967 \\ 
 1913$-$0440 &    1911$-$04 & 19:13:54.18(1)~ &  $-$04:40:47.6(4)~ &     9 & 19:13:54.184(5)~~~ & $-$04:40:47.6(3)~~ &    51550--51966 \\ 
   1917+1353 &      1915+13 & 19:17:39.784(2) &  +13:53:57.06(9)~~ &     2 & 19:17:39.790(4)~~~ & +13:53:56.80(9)~~~ &    51573--51966 \\ 
   1921+2153 &      1919+21 & 19:21:44.798(3) &  +21:53:01.83(8)~~ &     9 & 19:21:44.80(3)~~~~ & +21:53:02.8(5)~~~~ &    51600--51966 \\ 
   1932+1059 &      1929+10 & 19:32:13.900(2) &  +10:59:31.99(7)~~ &     2 & 19:32:13.9430(15)~ & +10:59:32.52(6)~~~ &    51550--51960 \\ 
 \vspace{-2mm} \\ 
   1935+1616 &      1933+16 & 19:35:47.835(1) &  +16:16:40.59(2)~~ &     1 & 19:35:47.8249(11)~ & +16:16:40.03(4)~~~ &    51550--51966 \\ 
   1946+1805 &      1944+17 & 19:46:53.043(4) &  +18:05:41.59(9)~~ &    17 & 19:46:53.05(3)~~~~ & +18:05:41.5(8)~~~~ &    51560--51966 \\ 
   1948+3540 &      1946+35 & 19:48:25.037(2) &  +35:40:11.28(2)~~ &    17 & 19:48:25.006(3)~~~ & +35:40:11.03(6)~~~ &    51500--51966 \\ 
   1955+5059 &      1953+50 & 19:55:18.90(6)~ &  +50:59:54.2(6)~~~ &    16 & 19:55:18.736(6)~~~ & +50:59:55.76(10)~~ &    51513--51966 \\ 
   2002+4050 &      2000+40 & 20:02:44.04(3)~ &  +40:50:54.7(3)~~~ &     5 & 20:02:44.047(10)~~ & +40:50:53.99(19)~~ &    51560--51966 \\ 
 \vspace{-2mm} \\ 
   2013+3845 &      2011+38 & 20:13:10.49(3)~ &  +38:45:44.8(3)~~~ &     5 & 20:13:10.384(7)~~~ & +38:45:43.08(16)~~ &    51549--51966 \\ 
   2018+2839 &      2016+28 & 20:18:03.851(2) &  +28:39:54.26(3)~~ &     1 & 20:18:03.832(4)~~~ & +28:39:54.26(9)~~~ &    51549--51966 \\ 
   2022+2854 &      2020+28 & 20:22:37.079(3) &  +28:54:23.45(3)~~ &    18 & 20:22:37.078(3)~~~ & +28:54:22.91(6)~~~ &    51547--51967 \\ 
   2022+5154 &      2021+51 & 20:22:49.900(2) &  +51:54:50.06(2)~~ &     1 & 20:22:49.843(10)~~ & +51:54:50.46(16)~~ &    51547--51968 \\ 
 2048$-$1616 &    2045$-$16 & 20:48:35.472(4) &  $-$16:16:44.45(8) &     3 & 20:48:35.46(8)~~~~ & $-$16:16:38(4)~~~~ &    51547--51960 \\ 
 \vspace{-2mm} \\ 
   2108+4441 &      2106+44 & 21:08:20.48(1)~ &  +44:41:48.8(1)~~~ &     2 & 21:08:20.521(8)~~~ & +44:41:49.37(16)~~ &    51547--51966 \\ 
   2113+4644 &      2111+46 & 21:13:24.29(1)~ &  +46:44:08.7(1)~~~ &     2 & 21:13:24.377(10)~~ & +46:44:09.1(3)~~~~ &    51547--51966 \\ 
   2157+4017 &      2154+40 & 21:57:01.82(1)~ &  +40:17:45.9(1)~~~ &     2 & 21:57:01.879(13)~~ & +40:17:45.8(4)~~~~ &    51547--51966 \\ 
   2219+4754 &      2217+47 & 22:19:48.136(4) &  +47:54:53.83(4)~~ &     2 & 22:19:48.115(10)~~ & +47:54:53.73(12)~~ &    51506--51966 \\ 
   2257+5909 &      2255+58 & 22:57:57.711(4) &  +59:09:14.95(3)~~ &     2 & 22:57:57.756(9)~~~ & +59:09:14.93(9)~~~ &    51547--51966 \\ 
 \vspace{-2mm} \\ 
   2313+4253 &      2310+42 & 23:13:08.571(6) &  +42:53:12.98(3)~~ &     3 & 23:13:08.622(10)~~ & +42:53:13.12(19)~~ &    51547--51967 \\ 
   2321+6024 &      2319+60 & 23:21:55.19(4)~ &  +60:24:30.7(3)~~~ &     2 & 23:21:55.16(3)~~~~ & +60:24:30.58(16)~~ &    51499--51966 \\ 
   2326+6113 &      2324+60 & 23:26:58.704(5) &  +61:13:36.50(3)~~ &     2 & 23:26:58.693(8)~~~ & +61:13:36.29(8)~~~ &    51560--51966 \\ 
   2354+6155 &      2351+61 & 23:54:04.71(2)~ &  +61:55:46.8(1)~~~ &     2 & 23:54:04.743(16)~~ & +61:55:46.67(12)~~ &    51547--51966 \\ 
\hline&\vspace{-8mm}
\end{tabular}
\label{tb:position2}
\end{minipage}
\end{table*}

%% file: rotation1.tex
\begin{table*}
\begin{minipage}{150mm}
\caption{Pulsar period parameters at MJD 51700.}
\begin{tabular}{ccllclllrc}
\hline &\vspace{-3mm} \\
PSR J     & PSR B         & ~~~~~~~~~$P$     & ~~~~~~~~~$\dot P$
& $N_{\rm toa}$&  Res.    & ~~~~~~~~~~$\Delta P$   & ~~~~~~$\Delta \dot P$ 
& !$\Delta T$     & 
 Ref.\footnote{
   1. Manchester \& Taylor, 1981 \nocite{mt81} 
~~~2. Arzoumanian et al., 1994 \nocite{antt94}
~~~3. Backus, Taylor \& Damashek, 1982 \nocite{btd82}
~~~4. Shemar \& Lyne, 1996 \nocite{sl96}
~~~5. Siegman, Manchester \& Durdin, 1993 \nocite{smd93}
~~~6. Dewey et al., 1988 \nocite{dtms88}
~~~7. Manchester et al. 1983 \nocite{mnhg83}
~~~8. Newton, Manchester \& Cooke, 1981 \nocite{nmc81} 
~~~9. Gullahorn \& Rankin, 1978 \nocite{gr78}
~~~10. Vivekanand, Mohanty \& Salter, 1983 \nocite{vms83}
~~~11. Wang et al., 2000 \nocite{wmp+00}
~~~12. Clifton et al., 1992 \nocite{clj+92}
~~~13. Manchester et al., 1996 \nocite{mld+96}
~~~14. Foster, Backer \& Wolszczan, 1990 \nocite{fbw90}
}      \\
             &                 &  ~~~~~~~~(s)      & 
~~~~~($10^{-15}$)   &  & ($\mu$s) & ~~~~~~($10^{-9}$ s)   &
~~~($10^{-18}$)  & (yr)         &                 \\
\hline &\vspace{-3mm} \\
0034$-$0721  & 0031$-$07    &  0.94295117325(7)~~~ &   !!!0.395(13)~~ &     25 &    854 &!!!$-$0.03(87)~ &  !$-$13(13)~~~ &   30.1 &      1 \\ 
0139+5814    & 0136+57      &  0.272452862190(3)~~ &   !!10.7040(6)~~ &     60 &    268 &!!!!~3.07(9)~~~ &  !!!~3.4(6)~~~ &    9.1 &      2 \\ 
0141+6009    & 0138+59      &  1.22294860175(7)~~~ &   !!!0.417(17)~~ &     41 &    756 &!!!!~0.4(7)~~~~ &  !!~26(17)~~~~ &   27.1 &      3 \\ 
0332+5434    & 0329+54      &  0.714520624711(5)~~ &   !!!2.0489(10)~ &    102 &    469 &!!!$-$1.19(8)~~ &  !!$-$0.7(10)~ &   30.3 &      1 \\ 
0358+5413    & 0355+54      &  0.1563832093117(10) &   !!!4.3954(3)~~ &     61 &    342 &!!!!~0.66(10)~~ &  !!!~5.0(5)~~~ &   14.2 &      4 \\ 
  \vspace{-1mm} \\ 
0452$-$1759  & 0450$-$18    &  0.548940421758(19)~ &   !!!5.753(3)~~~ &     54 &    546 &!!!$-$1.31(18)~ &  !!$-$4(3)~~~~ &   13.4 &      5 \\ 
0454+5543    & 0450+55      &  0.340729803314(5)~~ &   !!!2.3742(8)~~ &     59 &    321 &!!!!~1.95(6)~~~ &  !!!~8.6(8)~~~ &    9.0 &      2 \\ 
0528+2200    & 0525+21      &  3.74553060709(15)~~ &   !!40.01(3)~~~~ &     52 &    893 &!!!!~2(11)~~~~~ &  !$-$16(36)~~~ &   26.4 &      4 \\ 
0543+2329    & 0540+23      &  0.245978510226(3)~~ &   !!15.4205(7)~~ &     65 &    568 &!!!$-$0.764(18) &  !!$-$3.8(7)~~ &    9.1 &      2 \\ 
0612+3721    & 0609+37      &  0.297982336991(18)~ &   !!!0.059(3)~~~ &     21 &    293 &!!!!~0.23(6)~~~ &  !!!~0.05(300) &   15.3 &      6 \\ 
  \vspace{-1mm} \\ 
0630$-$2834  & 0628$-$28    &  1.24442173630(8)~~~ &   !!!7.195(14)~~ &     53 &   1171 &!!!~11(6)~~~~~~ &  !!~88(16)~~~~ &   20.7 &      7 \\ 
0738$-$4042  & 0736$-$40    &  0.374919985032(6)~~ &   !!!1.6161(9)~~ &     44 &    171 &!!!$-$1.3(3)~~~ &  !!!~2.0(9)~~~ &   25.0 &      7 \\ 
0742$-$2822  & 0740$-$28    &  0.166765742577(5)~~ &   !!16.8177(8)~~ &     62 &    575 &!!!!~3.20(6)~~~ &  !!!~5.2(8)~~~ &    9.1 &      2 \\ 
0814+7429    & 0809+74      &  1.29224148381(11)~~ &   !!!0.153(16)~~ &     57 &   1119 &!!!!~0.03(26)~~ &  !$-$15(16)~~~ &    9.1 &      2 \\ 
0820$-$1350  & 0818$-$13    &  1.23813005246(4)~~~ &   !!!2.097(6)~~~ &     58 &    644 &!!!$-$0.24(19)~ &  !!$-$8(6)~~~~ &   29.3 &      1 \\ 
  \vspace{-1mm} \\ 
0826+2637    & 0823+26      &  0.530661285865(12)~ &   !!!1.6978(19)~ &     66 &    588 &!!!$-$1.61(12)~ &  !$-$11.6(20)~ &    9.1 &      2 \\ 
0837$-$4135  & 0835$-$41    &  0.751623617646(8)~~ &   !!!3.5393(12)~ &     33 &    165 &!!!$-$1.8(3)~~~ &  !!$-$7.6(13)~ &   13.4 &      5 \\ 
0846$-$3533  & 0844$-$35    &  1.1160975763(3)~~~~ &   !!!1.62(5)~~~~ &     30 &   1046 &!!!~22(22)~~~~~ &  !!~48(53)~~~~ &   22.3 &      8 \\ 
0922+0638    & 0919+06      &  0.430623569113(12)~ &   !!13.6924(19)~ &     54 &    521 &!!!$-$1.18(15)~ &  !$-$28.1(20)~ &    9.0 &      2 \\ 
0953+0755    & 0950+08      &  0.253065270746(6)~~ &   !!!0.2288(10)~ &     69 &    540 &!!!!~0.62(10)~~ &  !!$-$0.4(10)~ &   27.9 &      9 \\ 
  \vspace{-1mm} \\ 
1136+1551    & 1133+16      &  1.18791477322(3)~~~ &   !!!3.733(5)~~~ &     73 &    661 &!!!!~0.60(7)~~~ &  !!!~0.6(48)~~ &   27.5 &      9 \\ 
1239+2453    & 1237+25      &  1.38244953180(4)~~~ &   !!!0.967(7)~~~ &     67 &    677 &!!!$-$0.15(15)~ &  !!!~6(7)~~~~~ &    9.1 &      2 \\ 
1509+5531    & 1508+55      &  0.739682698169(16)~ &   !!!4.996(3)~~~ &     58 &    512 &!!!$-$2.68(9)~~ &  !$-$12(3)~~~~ &    9.1 &      2 \\ 
1645$-$0317  & 1642$-$03    &  0.387690495052(10)~ &   !!!1.7924(15)~ &     67 &    595 &!!!$-$1.04(10)~ &  !!~11.4(15)~~ &   30.3 &      1 \\ 
1703$-$3241  & 1700$-$32    &  1.21178519124(4)~~~ &   !!!0.660(7)~~~ &     57 &    709 &!!$-$36(31)~~~~ &  $-$130(97)~~~ &   10.1 &      2 \\ 
  \vspace{-1mm} \\ 
1705$-$1906  & 1702$-$19    &  0.298988488005(8)~~ &   !!!4.1370(14)~ &     55 &    603 &!!!$-$0.244(20) &  !!$-$1.3(14)~ &    9.2 &      2 \\ 
1707$-$4053  & 1703$-$40    &  0.581016707860(18)~ &   !!!1.925(4)~~~ &     23 &    280 &!!!!~0.19(7)~~~ &  !!!~5(4)~~~~~ &    9.1 &      2 \\ 
1709$-$1640  & 1706$-$16    &  0.65305652375(3)~~~ &   !!!6.240(4)~~~ &     50 &    675 &!!$-$11.99(12)~ &  !$-$69(4)~~~~ &    9.1 &      2 \\ 
1721$-$3532  & 1718$-$35    &  0.280424567967(9)~~ &   !!25.1816(14)~ &     52 &    497 &!!!$-$4.28(9)~~ &  !$-$14.5(15)~ &    9.1 &      2 \\ 
1722$-$3207  & 1718$-$32    &  0.477157619208(5)~~ &   !!!0.6475(9)~~ &     57 &    378 &!!!$-$0.25(4)~~ &  !!!~0.4(9)~~~ &    9.1 &      2 \\ 
  \vspace{-1mm} \\ 
1740$-$3015  & 1737$-$30    &  0.60676381008(3)~~~ &   !466.312(5)~~~ &     48 &    893 &$-$1181(9)~~~~~ &  !~291(54)~~~~ &    6.7 &      4 \\ 
1741$-$3927  & 1737$-$39    &  0.512211401067(8)~~ &   !!!1.9313(14)~ &     44 &    319 &!!!~68(3)~~~~~~ &  !~122(5)~~~~~ &   22.3 &      8 \\ 
1745$-$3040  & 1742$-$30    &  0.367430422265(6)~~ &   !!10.6696(10)~ &     54 &    327 &!!!!~0.86(3)~~~ &  !!!~4.5(10)~~ &    9.1 &      2 \\ 
1752$-$2806  & 1749$-$28    &  0.562561300009(9)~~ &   !!!8.1271(16)~ &     62 &    550 &!!!!~2.0(7)~~~~ &  !!!~5(3)~~~~~ &    9.5 &      5 \\ 
1757$-$2421  & 1754$-$24    &  0.234102570587(4)~~ &   !!12.9146(5)~~ &     51 &    373 &!!$-$54(192)~~~ &  !$-$100(300)~ &   20.2 &     10 \\ 
  \vspace{-1mm} \\ 
1803$-$2137  & 1800$-$21    &  0.133641979388(15)~ &   !134.150(3)~~~ &     36 &   2378 &!!!!~5(6)~~~~~~ &  !~206(200)~~~ &    1.9 &     11 \\ 
1807$-$0847  & 1804$-$08    &  0.163727380271(3)~~ &   !!!0.0287(5)~~ &     58 &    725 &!!!!~0.09(7)~~~ &  !!!~0.03(48)~ &   21.4 &      3 \\ 
1818$-$1422  & 1815$-$14    &  0.291489038008(5)~~ &   !!!2.0399(9)~~ &     49 &    316 &!!!!~0.9(3)~~~~ &  !!!~3.0(11)~~ &   12.9 &     12 \\ 
1820$-$0427  & 1818$-$04    &  0.598078699467(10)~ &   !!!6.3379(20)~ &     60 &    744 &!!!$-$6.1(5)~~~ &  !!!~0.2(21)~~ &   30.3 &      1 \\ 
1824$-$1945  & 1821$-$19    &  0.189335814049(3)~~ &   !!!5.2394(5)~~ &     63 &    519 &!!!!~2.3(6)~~~~ &  !!~14.5(14)~~ &   13.4 &      5 \\ 
  \vspace{-1mm} \\ 
1825$-$0935  & 1822$-$09    &  0.768994380569(16)~ &   !!52.336(4)~~~ &     59 &    909 &!!$-$33.5(3)~~~ &  !$-$29(4)~~~~ &    9.1 &      2 \\ 
1829$-$1751  & 1826$-$17    &  0.307133849665(11)~ &   !!!5.5414(16)~ &     50 &    420 &!!!$-$4.17(13)~ &  !$-$20.6(16)~ &   13.4 &      5 \\ 
1832$-$0827  & 1829$-$08    &  0.647305181037(9)~~ &   !!63.8874(20)~ &     55 &    713 &!!!$-$3.70(17)~ &  !!$-$5.8(20)~ &   12.9 &     12 \\ 
1833$-$0827  & 1830$-$08    &  0.0852852156141(15) &   !!!9.1721(3)~~ &     39 &    145 &!!!!~0.30(5)~~~ &  !!!~2.6(3)~~~ &   10.0 &      4 \\ 
1835$-$1106  &              &  0.165910996503(20)~ &   !!20.656(3)~~~ &     27 &    915 &!!!!~3.896(20)~ &  !!~44(3)~~~~~ &    6.2 &     13 \\ 
  \vspace{-1mm} \\ 
1836$-$1008  & 1834$-$10    &  0.562713612837(7)~~ &   !!11.8013(14)~ &     58 &    618 &!!!~15(5)~~~~~~ &  !!~26(7)~~~~~ &   21.4 &      3 \\ 
1840+5640    & 1839+56      &  1.65286223816(14)~~ &   !!!1.51(3)~~~~ &     46 &   1812 &!!!$-$0.01(88)~ &  !!~10(24)~~~~ &    9.1 &      2 \\ 
1847$-$0402  & 1844$-$04    &  0.59778239230(3)~~~ &   !!51.704(4)~~~ &     50 &    708 &!!!$-$2.3(5)~~~ &  !!$-$10(4)~~~ &   13.4 &      5 \\ 
1848$-$0123  & 1845$-$01    &  0.659432812074(9)~~ &   !!!5.2479(18)~ &     58 &    709 &!!!~11.6(7)~~~~ &  !!~29(3)~~~~~ &   13.4 &      5 \\ 
1900$-$2600  & 1857$-$26    &  0.61220925419(4)~~~ &   !!!0.193(6)~~~ &     49 &    788 &!!!!~0.20(6)~~~ &  !$-$11(6)~~~~ &    9.1 &      2 \\ 
\hline&\vspace{-8mm} 
\end{tabular}
\label{tb:rotation1}
\end{minipage}
\end{table*}

%% file: rotation2.tex
\begin{table*}
\begin{minipage}{150mm}
\caption{Pulsar period parameters at MJD 51700.}
\begin{tabular}{ccllclllrc}
\hline &\vspace{-3mm} \\
PSR J     & PSR B         & ~~~~~~~~~$P$     & ~~~~~~~~~$\dot P$
& $N_{\rm toa}$&  Res.    & ~~~~~~~~~~$\Delta P$   & ~~~~~~$\Delta \dot P$ 
& !$\Delta T$     & 
 Ref.\footnote{
   1. Manchester \& Taylor, 1981 \nocite{mt81} 
~~~2. Arzoumanian et al., 1994 \nocite{antt94}
~~~3. Backus, Taylor \& Damashek, 1982 \nocite{btd82}
~~~4. Shemar \& Lyne, 1996 \nocite{sl96}
~~~5. Siegman, Manchester \& Durdin, 1993 \nocite{smd93}
~~~6. Dewey et al., 1988 \nocite{dtms88}
~~~7. Manchester et al. 1983 \nocite{mnhg83}
~~~8. Newton, Manchester \& Cooke, 1981 \nocite{nmc81} 
~~~9. Gullahorn \& Rankin, 1978 \nocite{gr78}
~~~10. Vivekanand, Mohanty \& Salter, 1983 \nocite{vms83}
~~~11. Wang et al., 2000 \nocite{wmp+00}
~~~12. Clifton et al., 1992 \nocite{clj+92}
~~~13. Manchester et al., 1996 \nocite{mld+96}
~~~14. Foster, Backer \& Wolszczan, 1990 \nocite{fbw90}
}      \\
             &                 &  ~~~~~~~~(s)      & 
~~~~~($10^{-15}$)   &  & ($\mu$s) & ~~~~~~($10^{-9}$ s)   &
~~~($10^{-18}$)  & (yr)         &                 \\
\hline &\vspace{-3mm} \\
1913$-$0440  & 1911$-$04    &  0.825937582927(13)~ &   !!!4.055(3)~~~ &     47 &    301 &!!!$-$1.37(20)~ &  !$-$14(3)~~~~ &   30.3 &      1 \\ 
1917+1353    & 1915+13      &  0.1946321872566(13) &   !!!7.19502(18) &     49 &    256 &!!!$-$0.150(18) &  !!$-$3.22(19) &    9.1 &      2 \\ 
1921+2153    & 1919+21      &  1.33730247470(8)~~~ &   !!!1.358(8)~~~ &     26 &   1151 &!!!$-$0.02(9)~~ &  !!!~10(8)~~~~ &   30.1 &      1 \\ 
1932+1059    & 1929+10      &  0.2265181529285(15) &   !!!1.1618(3)~~ &     39 &    120 &!!!!~0.39(4)~~~ &  !!!~5.2(3)~~~ &    9.1 &      2 \\ 
1935+1616    & 1933+16      &  0.3587411416716(16) &   !!!6.0023(3)~~ &     42 &     84 &!!!$-$0.91(5)~~ &  !!$-$1.4(3)~~ &   25.8 &      9 \\ 
  \vspace{-1mm} \\ 
1946+1805    & 1944+17      &  0.44061848299(3)~~~ &   !!!0.020(3)~~~ &     54 &   2234 &!!!$-$0.07(7)~~ &  !!$-$4(3)~~~~ &    9.1 &      2 \\ 
1948+3540    & 1946+35      &  0.717312547254(5)~~ &   !!!7.0592(10)~ &     54 &    340 &!!!!~6.06(8)~~~ &  !!!~7.0(10)~~ &   26.0 &      9 \\ 
1955+5059    & 1953+50      &  0.518938338380(8)~~ &   !!!1.3722(15)~ &     59 &    539 &!!!!~4.0(14)~~~ &  !!!~6(3)~~~~~ &   21.4 &      3 \\ 
2002+4050    & 2000+40      &  0.90506709268(4)~~~ &   !!!1.743(6)~~~ &     53 &    757 &!!!$-$2.6(10)~~ &  !!$-$0.9(62)~ &   15.3 &      6 \\ 
2013+3845    & 2011+38      &  0.230195129550(7)~~ &   !!!8.8514(11)~ &     50 &    602 &!!!$-$1.52(20)~ &  !!$-$3.7(12)~ &   15.4 &      6 \\ 
  \vspace{-1mm} \\ 
2018+2839    & 2016+28      &  0.557953548298(8)~~ &   !!!0.1478(13)~ &     42 &    282 &!!!$-$1.08(3)~~ &  !!$-$1.6(13)~ &   30.1 &      1 \\ 
2022+2854    & 2020+28      &  0.343402486392(4)~~ &   !!!1.8924(6)~~ &     44 &    203 &!!!!~0.3(3)~~~~ &  !!$-$1.1(8)~~ &   12.8 &     14 \\ 
2022+5154    & 2021+51      &  0.529198257306(18)~ &   !!!3.061(3)~~~ &     72 &    884 &!!!$-$0.73(4)~~ &  !!$-$4(3)~~~~ &    9.1 &      2 \\ 
2048$-$1616  & 2045$-$16    &  1.96157730029(5)~~~ &   !!10.973(9)~~~ &     41 &    474 &!!!$-$2.15(20)~ &  !!~12(9)~~~~~ &   30.1 &      1 \\ 
2108+4441    & 2106+44      &  0.414870558915(12)~ &   !!!0.095(3)~~~ &     52 &    679 &!!!$-$0.02(5)~~ &  !!!~9(3)~~~~~ &    9.1 &      2 \\ 
  \vspace{-1mm} \\ 
2113+4644    & 2111+46      &  1.01468510534(4)~~~ &   !!!0.719(7)~~~ &     48 &    834 &!!!$-$0.84(18)~ &  !!!~7(7)~~~~~ &    9.1 &      2 \\ 
2157+4017    & 2154+40      &  1.52526635476(9)~~~ &   !!!3.434(14)~~ &     61 &   1308 &!!!!~4.0(3)~~~~ &  !!!~8(14)~~~~ &    9.1 &      2 \\ 
2219+4754    & 2217+47      &  0.538470040747(10)~ &   !!!2.7634(20)~ &     57 &    800 &!!!!~0.13(3)~~~ &  !!$-$1.6(20)~ &    9.1 &      2 \\ 
2257+5909    & 2255+58      &  0.368247257217(8)~~ &   !!!5.7555(13)~ &     61 &    490 &!!!$-$0.19(5)~~ &  !!!~1.7(13)~~ &    9.0 &      2 \\ 
2313+4253    & 2310+42      &  0.349433715696(12)~ &   !!!0.111(3)~~~ &     58 &    819 &!!!$-$2.0(6)~~~ &  !!$-$4(3)~~~~ &   21.4 &      3 \\ 
  \vspace{-1mm} \\ 
2321+6024    & 2319+60      &  2.25648988440(7)~~~ &   !!!7.017(13)~~ &     58 &   1322 &!!!!~0.3(6)~~~~ &  !$-$20(13)~~~ &    9.1 &      2 \\ 
2326+6113    & 2324+60      &  0.233652031097(5)~~ &   !!!0.3513(8)~~ &     53 &    360 &!!!$-$0.008(10) &  !!$-$1.3(8)~~ &    9.0 &      2 \\ 
2354+6155    & 2351+61      &  0.94478711093(3)~~~ &   !!16.248(5)~~~ &     56 &    685 &!!!$-$0.59(12)~ &  !$-$16(5)~~~~ &    9.1 &      2 \\ 
\hline&\vspace{-8mm} 
\end{tabular}
\end{minipage}
\end{table*}